\documentclass[twocolumn]{aa}
\usepackage{graphicx}
\usepackage{txfonts}
\usepackage[authoryear]{natbib}
\bibpunct{(}{)}{;}{a}{}{,}
\newcommand{\igr}  {IGR J06074+2205}
\newcommand{\ha}  {H$\alpha$}
\newcommand{\ew}  {EW(H$\alpha$)}
\newcommand{\ergs}  {erg s$^{-1}$}

\def\simless{\mathbin{\lower 3pt\hbox
     {$\rlap{\raise 5pt\hbox{$\char'074$}}\mathchar"7218$}}}   
\def\simmore{\mathbin{\lower 3pt\hbox
     {$\rlap{\raise 5pt\hbox{$\char'076$}}\mathchar"7218$}}}   

\def\msun{~{\rm M}_\odot}

\begin{document}

   \title{Discovery of X-ray pulsations in the Be/X-ray binary \\ IGR J06074+2205 }

   \subtitle{}
  \author{
	P. Reig\inst{1,2}
	\and
  	A. Zezas\inst{2,3}
           }

\authorrunning{Reig et al.}
\titlerunning{Discovery of X-ray pulsation in \igr}

   \offprints{pau@physics.uoc.gr}

   \institute{IESL, Foundation for Reseach and Technology-Hellas, 71110, 
   		Heraklion, Greece 
	 \and Physics Department, University of Crete, 71003, 
   		Heraklion, Greece 
		\email{pau@physics.uoc.gr}
	 \and Harvard-Smithsonian Center for Astrophysics, 60 Garden
	 Street, Cambridge, MA02138, USA
	}

   \date{Received ; accepted}

\abstract
{IGR J06074+2205 is a poorly studied X-ray source with a Be
star companion. It has been proposed to belong to the group of Be/X-ray
binaries.  In Be/X-ray 
binaries, accretion onto the neutron star occurs via the transfer of material 
from the Be star's circumstellar disk. Thus, in the absence of the disk, 
no X-ray should be detected. }
{The main goal of this work is to study the quiescent X-ray emission of IGR
J06074+2205 during a disk-loss episode.  } 
{We obtained light curves at different energy bands and a spectrum covering the
energy range 0.4--12 keV. We used Fourier analysis to study the aperiodic
variability and epoch folding methods to study the periodic variability. Model
fitting to the energy spectrum allowed us to identify the possible physical
processes that generated the X-rays. 
}
{We show that at the time of the {{\it XMM-Newton}} observation the decretion
disk around the Be star had vanished. Still, accretion appears as the source of
energy that powers the high-energy radiation in IGR J06074+2205. We report the
discovery of X-ray pulsations with a pulse period of 373.2 s and a pulse
fraction  of $\sim$50\%.   The 0.4--12 keV spectrum is well described  by an
absorbed power  law and blackbody components with the best fitting parameters: 
$N_{\rm H}=(6.2\pm0.5) \times 10^{21}$ cm$^{-2}$, $kT_{\rm bb}=1.16\pm0.03$ keV,
and $\Gamma=1.5\pm0.1$ The absorbed X-ray luminosity is $L_{\rm X}=1.4 \times 10^{34}$
erg s$^{-1}$ assuming a distance of 4.5 kpc.
}
{The detection of X-ray pulsations confirms the nature of IGR J06074+2205 as a
Be/X-ray binary. We discuss various scenarios to explain the quiescent X-ray
emission of this pulsar. We rule out cooling of the neutron star surface and
magnetospheric emission and conclude that accretion is the most likely scenario.
The origin of the accreted material remains an open question.  }


   \maketitle

\section{Introduction}

IGRJ06074+2205 was discovered by {\it INTEGRAL}/JEM-X during public observations
of the Crab region that took place on 15 and 16 February 2003
\citep{chenevez04}. The source was detected with a flux of $\sim$ 22 mCrab
($\pm$2 mCrab) in the energy range 3--20 keV.  To localize the X-ray source and
determine the correct counterpart, \citet{tomsick06} obtained a 5 ks observation
with the Chandra X-ray Observatory on 2 December 2006. The {\it Chandra} flux
was $2 \times 10^{-12}$ erg cm$^{-2}$ s$^{-1}$ in the energy range 0.3-10 keV.
The energy spectrum could be fitted with an absorbed power-law with $N_H=
(6\pm2) \times 10^{22}$ cm$^{-2}$ and a photon index  of 1.3 $\pm$ 0.8 (90\%
confidence errors). The absorbed flux  was nearly 60 times lower than the value
at the time of the {\it INTEGRAL} observation. The {\it Chandra} observation
also allowed the confirmation of the Be star suggested by  \citet{halpern05} as
the correct optical counterpart.

Optical spectroscopic observations performed between 2006 and 2010 revealed a
fast-rotating ($v \sin i\approx 260$) B0.5Ve star located at $\sim$ 4.5 kpc
\citep{reig10b}. The star displayed variability in the \ha\ line  on timescales
of months. Changes were seen both in the line profile and the line intensity. 
Even though X-ray pulsations were not detected, the available data suggested
that \igr\ is a BeXB. BeXBs constitute the most numerous subgroup of high-mass
X-ray binaries and consist of a Be star and a neutron star
\citep{paul11,reig11}.  The mass donor in these systems is a relatively massive
($\simmore 10 \msun$) and fast-rotating ($\simmore$80\% of break-up velocity)
star, whose equator is surrounded by a disk formed from photospheric plasma
ejected by the star.  \ha\ in emission  is typically the dominant feature in the
spectra of Be stars and the reason for the letter ``e" in the spectral type
identification. In addition to emission lines, Be stars also show variable
polarized light and infrared excess, i.e., more near-IR emission than a normal B
type star of the same spectral type. The disk forms and dissipates on timescales
of years. Because the physical conditions in the disk change substantially
during the formation and dissipation phases, so do the \ha\ intensity and the 
optical and infrared photometric colours. Thus the disk is responsible for the
long-term optical and infrared variability of the system.  Also, since the
main source of matter available for accretion is the circumstellar disk, also
known as decretion  disk, the X-ray variability is expected to be closely linked
to the state of the disk. In particular, when the disk disappears completely,
the mass transfer toward the neutron star should halt and the X-ray emission
should cease. However, at least three systems have shown X-ray emission during
disk-loss phases: 1A\,0535+262 \citep{negueruela00}, IGR\,J21343+4738
\citep{reig14b}, and SAX\,J2103.5+4545 \citep{reig14c}. All three show X-ray
pulsations during these observations. In addition, many more sources have been
detected in a quiescent state at luminosities $L_{\rm X} \simless 10^{34}$
\ergs\citep{tsygankov17b}. The origin of the X-ray emission and of the accreted
material in this state remain unsolved.

In this work,  we present  an X-ray analysis during one of these disk-loss
episodes giving us the rare opportunity to study a BeXB in the quiescent phase
where the reservoir of accretion gas is exhausted. We report the discovery of
X-ray pulsations with a spin period of 373.2 s.

\section{Observations}

\subsection{X-ray band}

\igr\ was observed by {\it XMM-Newton} on 29 September 2017 during revolution
3261. The observation (ObsID 0794581301) started at 11:49 hr UT and lasted for
$\sim$40 ks. The {\it XMM-Newton} Observatory \citep{jansen01} includes three
1500 cm$^2$ X-ray telescopes each with an European Photon Imaging Camera (EPIC)
at the focus. Two of the EPIC imaging spectrometers use MOS CCDs
\citep{turner01} and one uses PN CCDs \citep{struder01}. Reflection Grating
Spectrometers \citep[][]{denherder01} are located behind two of the
telescopes while the 30-cm optical monitor (OM) instrument has its own
optical/UV telescope \citep{mason01}. Data were reduced using the XMM-Newton
Science Analysis System (SAS version 16.1).

Prior to the extraction of the data products, we processed the Observation Data
Files (ODFs) to obtain calibrated and concatenated event lists.   We filtered
the event-list file to extract PN events  with a pixel pattern in the range 0 to
4 (single and doubles) and  MOS events with patterns 0 to 12. For the spectral
analysis, we used the more strict selection criterion {\em FLAG=0}, which omits
parts of the detector area like border pixels and columns with higher offset.
The first 16240 seconds are affected by flaring particle background and were
discarded for the spectral analysis. For the timing analysis, we performed a
barycentric correction, i.e. the arrival time of the photons was shifted to the
barycentre of the solar system. To generate the light curves and spectrum, we
extracted events from a circular region with radius 40 arcsec.   This size
is the maximum allowed to avoid including the CCD edge. It includes more than
90\% of the encircled energy of 1.5 keV photons.  To select the
background region, we chose a  region of the same size $\sim2$ arcmin away from
the source.

The EPIC instruments were used in the {\em full frame} mode. In this mode, all
pixels of all CCDs are read out and thus the full field of view is covered. The
highest possible time resolution in this mode is 73.4 ms for PN and 2.6 s for
MOS. \igr\ is the brightest source detected in the field of view. The source
region is free of pile-up as demonstrated by the fact that the observed
distribution of counts as a function of the PI channel of single and double
events agrees with the expected one. Using the SAS task {\it epatplot} we find
that the 0.5 - 2.0 keV observed-to-model singles and doubles pattern fractions
ratios are consistent with 1.0 within statistical errors (0.992$\pm$0.027 and
1.011$\pm$0.038 for single and double events, respectively).

We mainly used data from the EPIC-PN camera because it is the instrument with
the highest effective area and the highest time resolution. Nevertheless, we
checked that we obtained consistent results with different cameras.

\subsection{Optical band}

We observed the source with the OM/{\it XMM-Newton} in the {\em fast}
mode. We extracted light curves with a total dutation of 2000 s for the bands
$U$, $B$, and $V$ and 8800 s for the UV bands $UVW1$ ($\lambda_{\rm eff}=291$
nm), $UVW2$ ($\lambda_{\rm eff}=212$ nm), and $UVM2$ ($\lambda_{\rm eff}=231$
nm). Timing analysis on each of these light curves did not reveal any
variability. The average magnitudes are: $U=12.43$, $B=13.04$, and $V=12.34$.

The spectroscopic observations were made with the 1.3 m telescope of the
Skinakas Observatory (Crete, Greece). We used  2048$\times$2048 13.5 $\mu$m
pixel ANDOR IKON and a 1302 l~mm$^{-1}$ grating, giving a nominal dispersion of
$\sim$0.94 \AA/pixel.  Spectra of comparison lamps were taken before each
exposure in order to account for small variations of the wavelength calibration
during the night. To ensure an homogeneous processing of the spectra, they were
normalized with respect to the local continuum, which  was rectified to unity by
employing a spline fit.  A disk-loss episode is identified when the \ha\ line
reverts from emission into absorption. In the absence of the disk, we expect to
observe the typical photospheric absorption line of an early-type star.
We report here spectroscopic observations taken close in time to the
{\it XMM-Netwon} observation. For a long-term optical
variability study of the source, see \citet{reig16a}.

\begin{figure}
\begin{center}
\includegraphics[width=0.8\columnwidth]{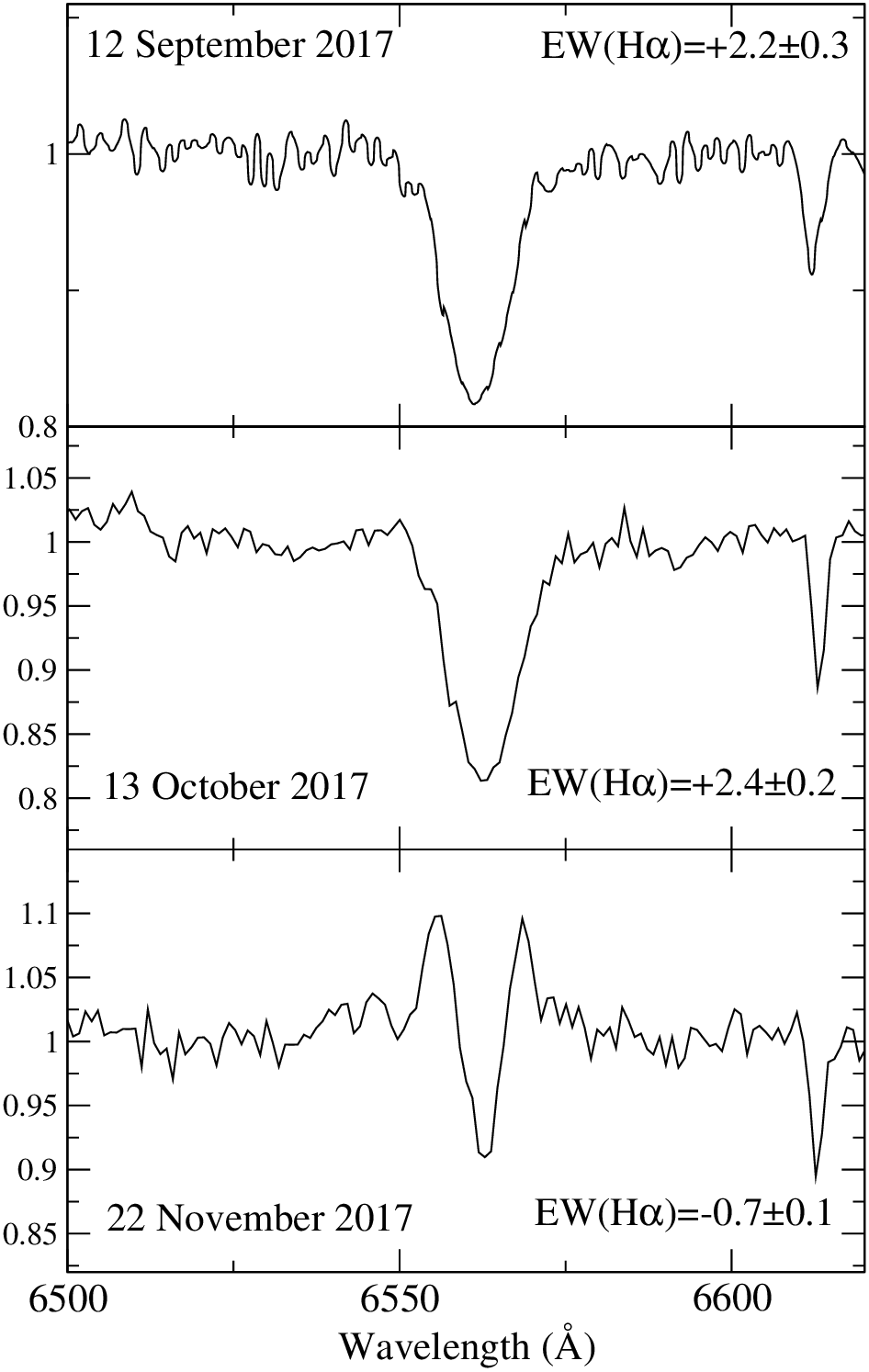}
\caption[]{\ha\ line profile before and after the X-ray observations on 29
September 2017. }
\label{haprof}
\end{center}
\end{figure}

\section{Results}

The optical/IR emission lines in Be stars arise from radiative processes in the
disk. In particular, the \ha\ line is optically thick and formed by
recombination.  It is the strongest feature and forms over a large region of the
disk at its outermost parts. Consequently, \ha\ emission
should disappear when the disk dissipates.  The most common way to quantify the
strength of the line is the equivalent width. Negative values indicate emission
profiles, while positive values correspond to absorption dominated profiles,
i.e., with the absence of the disk. Positive values of the \ew\ between 0 and
the expected value of a purely photospheric line, e.g. $\approx$+3.5 \AA\ for
B0.5V Be star \citep{jaschek87}, indicate that the line is partially filled with
emission.

Figure \ref{haprof} displays three \ha\ line profiles of \igr\ at
different epochs. The top and middle panels show the \ha\ line 17 days before
(12 September 2017) and 15 days after (13 October 2017) the {\it XMM-Newton}
observation. Both observations show \ha\ in absorption at a level of
\ew=+2.3\AA, indicating that the disk had vanished at the time of the X-ray
observation, although some weak residual emission cannot be completely ruled out.
In fact, the most recent observation, in November 2017 showed \ew$=-0.7$\AA\
(bottom panel in Fig.~\ref{haprof}), suggesting that the disk has begun to form again.

\begin{figure}
\resizebox{\hsize}{!}{\includegraphics{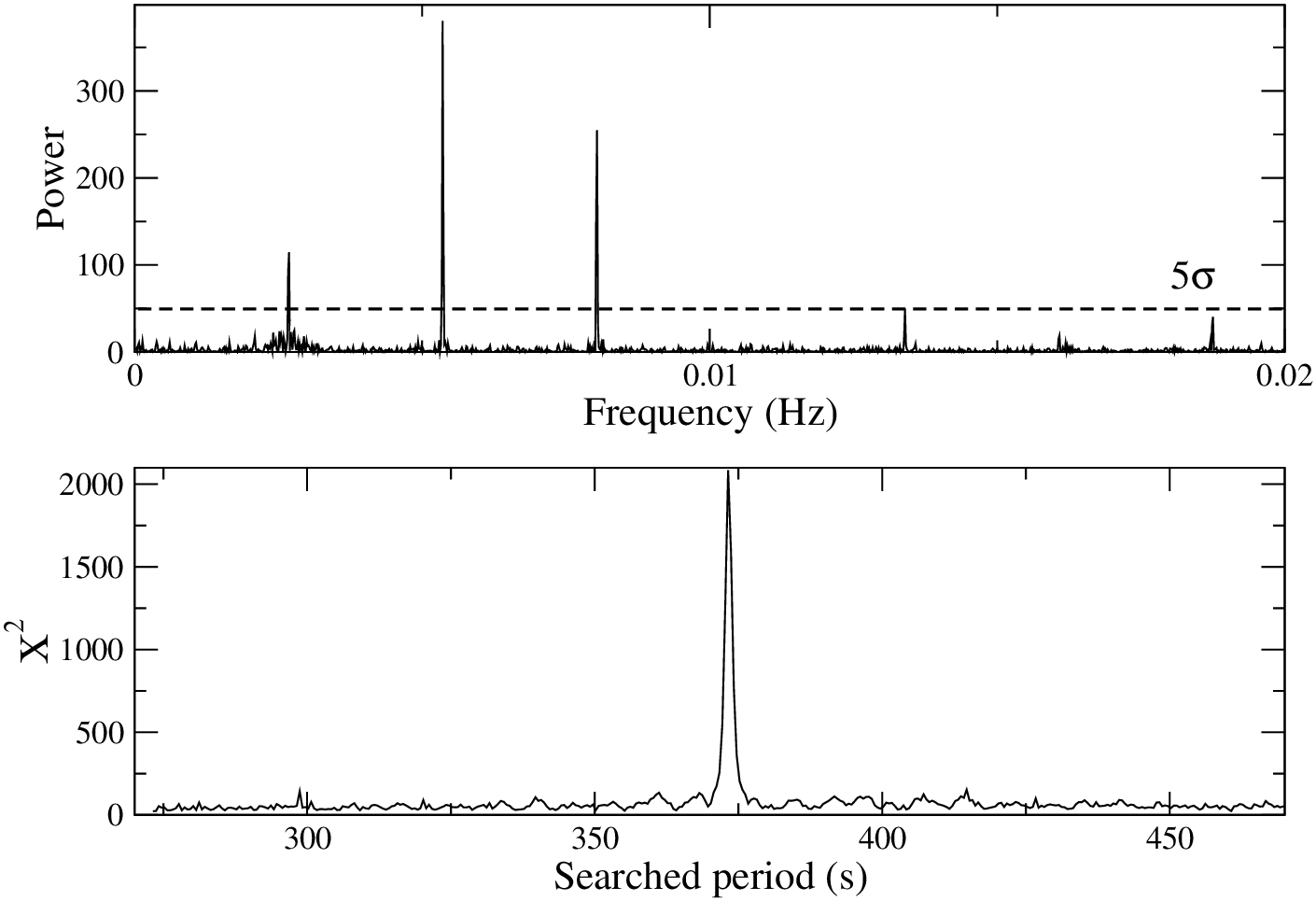} } 
\caption[]{{\em Upper panel}: the EPIC-PN power spectrum and $\sim5\sigma$
significance level. {\em Lower panel}: $\chi^2$ maximization after folding
the data over a range of periods (epoch folding).}
\label{spin}
\end{figure}

\subsection{Discovery of X-ray pulsations}
\label{per}

Although the material that feeds the neutron star was exhausted, X-ray emission
from \igr\ is clearly detected. To investigate the X-ray variability, we
extracted a 1 s binned light curve in the energy range  0.4--12 keV. 

The average 0.4--12 keV source count rate after background subtraction was
$0.970\pm0.006$ counts s$^{-1}$. The average background rate for the entire 38.8
ks observation was $0.036\pm0.001$ counts s$^{-1}$, but it was 
larger at the beginning of the observation owing to enhanced flaring particle
background. The average background count rate during the first 16 ks was
$0.062\pm0.002$ counts s$^{-1}$, while for the rest of the observation it
remained at $0.017\pm0.001$ counts s$^{-1}$.  The X-ray light curve exhibited
moderate variability on time scales of a few ks. The root-mean-square measured
in the 1 s binned light
curve was $43\pm1$\%.

To search for pulsations, we run a fast Fourier transform (FFT) and produce a 
Leahy-normalized \citep{leahy83} power spectrum covering the frequency interval
$3.05 \times 10^{-5}-0.5$ Hz in 32768 bins (Fig.~\ref{spin}). Three peaks are
clearly seen in the power spectrum at $\nu_1=0.002679$ Hz, $\nu_2=0.005356$ Hz,
and $\nu_3=0.008039$ Hz. All peaks are significant above 5~$\sigma$ level,
indicating that there is a distinct periodic signal in the data.  Because there
is a strong red noise component (see Sect.~\ref{pow}), the power spectrum was
normalized to unity and the result multiplied by 2. In this way we ensure that
the power spectral continuum agrees with the expected white noise power in the
Leahy normalization \citep{leahy83,leahy87}. This is a necessary step prior to
the calculation of the significant level. 

The power of the lowest frequency peak is distributed over several bins, making
this peak appear less significant than the second peak. However, the ratios of
the peak frequencies imply that the second peak cannot represent the fundamental
frequency. If $\nu_2$ was the fundamental frequency then $\nu_3$ would be the
first harmonic, hence  $\nu_3/\nu_2 = 2$. Clearly, $\nu_3/\nu_2 \neq 2$. In
contrast, if $\nu_1$ is the main periodicity, then $\nu_2/\nu_1=2$ and
$\nu_3/\nu_1=3$ as indeed it is the case. Therefore $\nu_2$ and $\nu_3$ are the
first and second harmonics of the main signal $\nu_1$ at 0.002685 Hz. This
frequency corresponds to a period of $\sim$372.4 s.

We also searched for periodic signal using an epoch-folding technique where the
data is folded over a period range \citep{leahy87,larsson96}. For each trial
period, the $\chi^2$ statistic is calculated. If the data contain a periodic
signal, then a peak stands out in the $\chi^2-P_{\rm trial}$ plot. We used the
task {\it efsearch} of the XRONOS package and found a best-fit period at 373.2 s
(Fig~\ref{spin}, lower panel). 

To improve the estimation of the pulse period, we next applied a phase fitting
technique. We divided the light curve into 8 segments, each of one with a length
equal to 12 pulse periods approximately and calculated a folded pulse profile
for each segment with a common epoch and period. These profiles were
cross-correlated with a template obtained by folding the entire light curve onto
the trial period. The resulting phase delays were fitted with a linear function,
whose slope provides the correction needed to be applied to the trial period. We
then adjusted the period and repeated the procedure until the phase delay
exhibited no net trend with time throughout the observation.  The best-fit
period was $373.226\pm0.013$~s, where the error was estimated from the
uncertainty on the first-order term of a linear fit to the phase delays.

\begin{figure}
\begin{center}
\includegraphics[width=0.8\columnwidth]{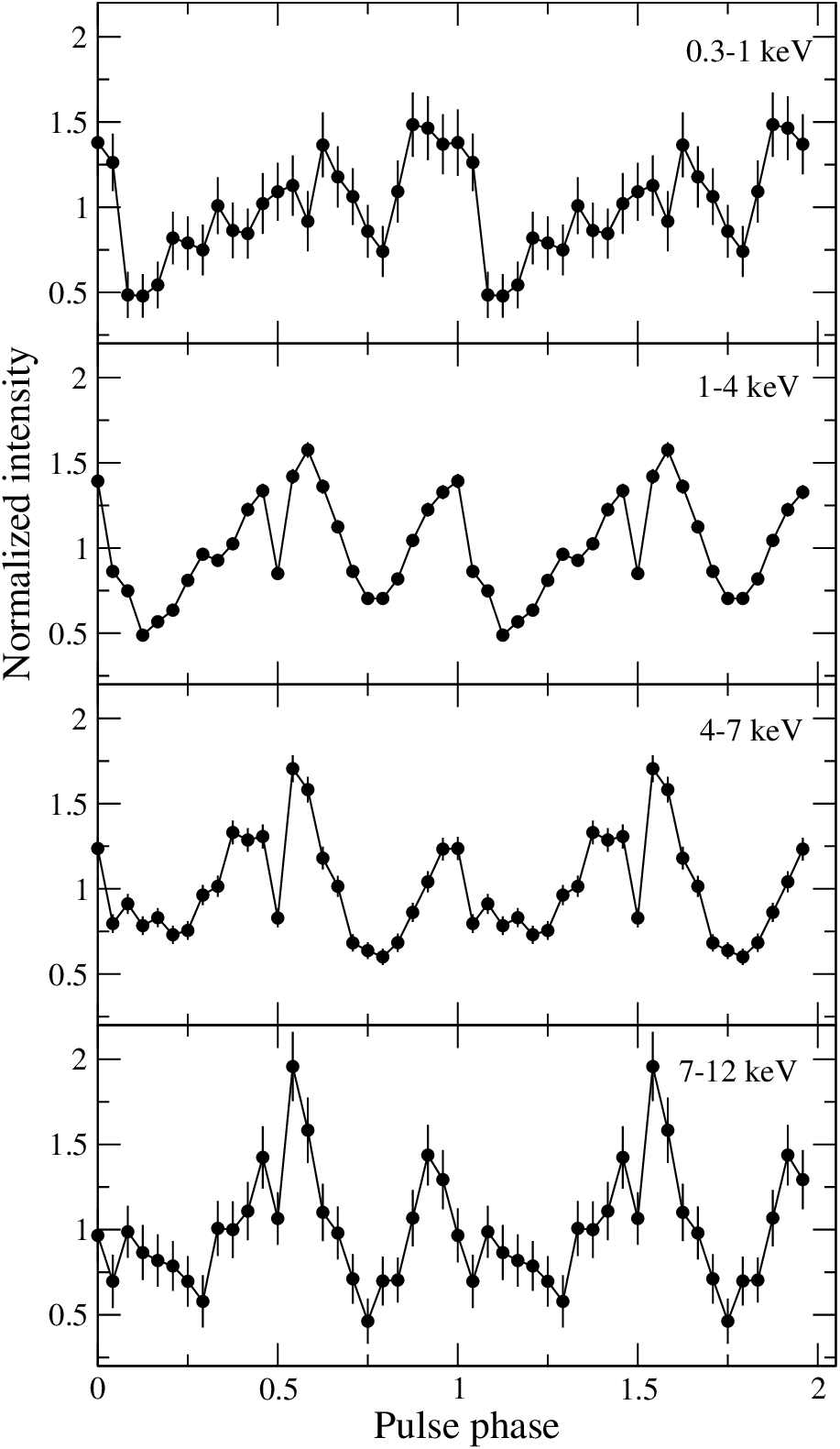}
\caption[]{Normalized pulse profiles at different energies.}
\label{profile}
\end{center}
\end{figure}

Next, we extracted light curves at various energy bands and folded the best-fit
pulse period onto these light curves to obtain the pulse profiles
(Fig.~\ref{profile}). The pulse profile is complex with a multi-peak structure.
The pulse fraction, defined as $PF = (I_{\rm max}-I_{\rm min})/(I_{\rm
max}+I_{\rm min})$, where $I_{\rm min}$ and $I_{\rm max}$ are
background-corrected count rates at the pulse profile minimum and maximum, does
not strongly depend on energy and varies in the range 50--60\%.

\begin{figure}
\resizebox{\hsize}{!}{\includegraphics{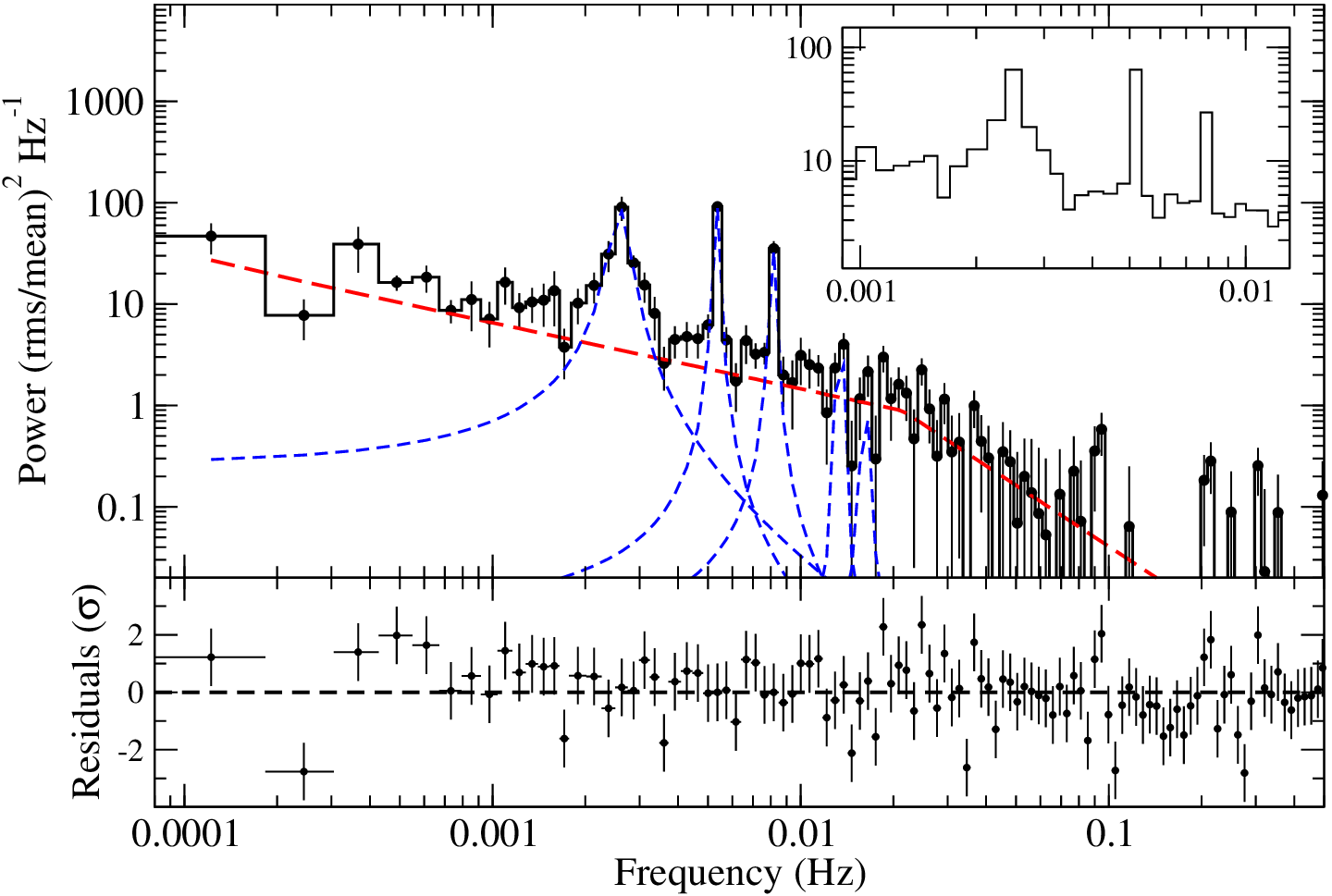} } 
\caption[]{Power spectrum of \igr\ (circles) and  model components, 
which consist of a broken power law (red dashed line) 
and multi Lorentzian profiles (blue lines).  The bottom panel gives the 
residuals in number of sigmas with error bars of size one. The inset shows with
detail the broadening at the base of the periodic signal.}
\label{powspec}
\end{figure}

\subsection{Power spectrum} 
\label{pow} 

The ability to detect a narrow feature in the power spectrum, i.e.
pulsations, is highly reduced when any kind of averaging (adjacent bins or
segments) is carried out \citep{klis89}. This is the reason that we used full
time resolution and no segment division in Sect.~\ref{per} and Fig.~\ref{spin}.
However, a study of the peaked and broad-band noise may also be of interest. In
this section, we examine the structure of the power spectrum to identify
possible noise components.  

To obtain the power spectrum shown in Fig.~\ref{powspec}, we proceeded as
follows: we extracted $\Delta t$ s binned light curves in the energy range
0.4--12 keV. The light curve was divided into segments of duration $T$ and a
Fast Fourier Transform was computed for each segment. The final power spectrum
is the average of all the power spectra obtained for each segment. The  power
spectra were logarithmically rebinned in frequency and normalized such that the
the integral between two frequencies gives the squared rms fractional
variability. The expected white noise level was subtracted. We tried different
combinations of $\Delta t$ and $T$ and found that above $\nu\sim0.1$ Hz the
power spectrum is white noise dominated. Therefore, we used $\Delta t=1$s and
$T=8192$ s covering a frequency range $0.00012-0.5$ Hz.

The power spectrum shows strong red noise. The spectral continuum can be
described by two zero-centred Lorentzians with widths $0.0019\pm0.0006$ Hz and
$0.023\pm0.006$ Hz, respectively (reduced $\chi^2=1.28$ for 98 degrees of
freedom) or with a broken power law with indices $\Gamma_1=0.65\pm0.05$ and
$\Gamma_2$ fixed to 2, and a break at $\nu_{\rm break}=0.024\pm0.004$ Hz
(reduced $\chi^2=1.26$ for 98 degrees of freedom).  The spikes caused by the
pulse period and its harmonics were fitted with Lorentzian profiles fixed at the
expected frequencies. The rms in the frequency range $0.0001-0.1$ Hz is 40\%.

An interesting feature of the power spectrum is the strong  coupling between the
periodic and aperiodic variability (see Sect.~\ref{coupling}), which manifests
as a broadening at the base of the harmonics \citep{lazzati97,burderi97} and
might be the reason that the power of the fundamental signal is smaller than
the harmonics (Fig.~\ref{spin}).

\begin{figure}
\resizebox{\hsize}{!}{\includegraphics{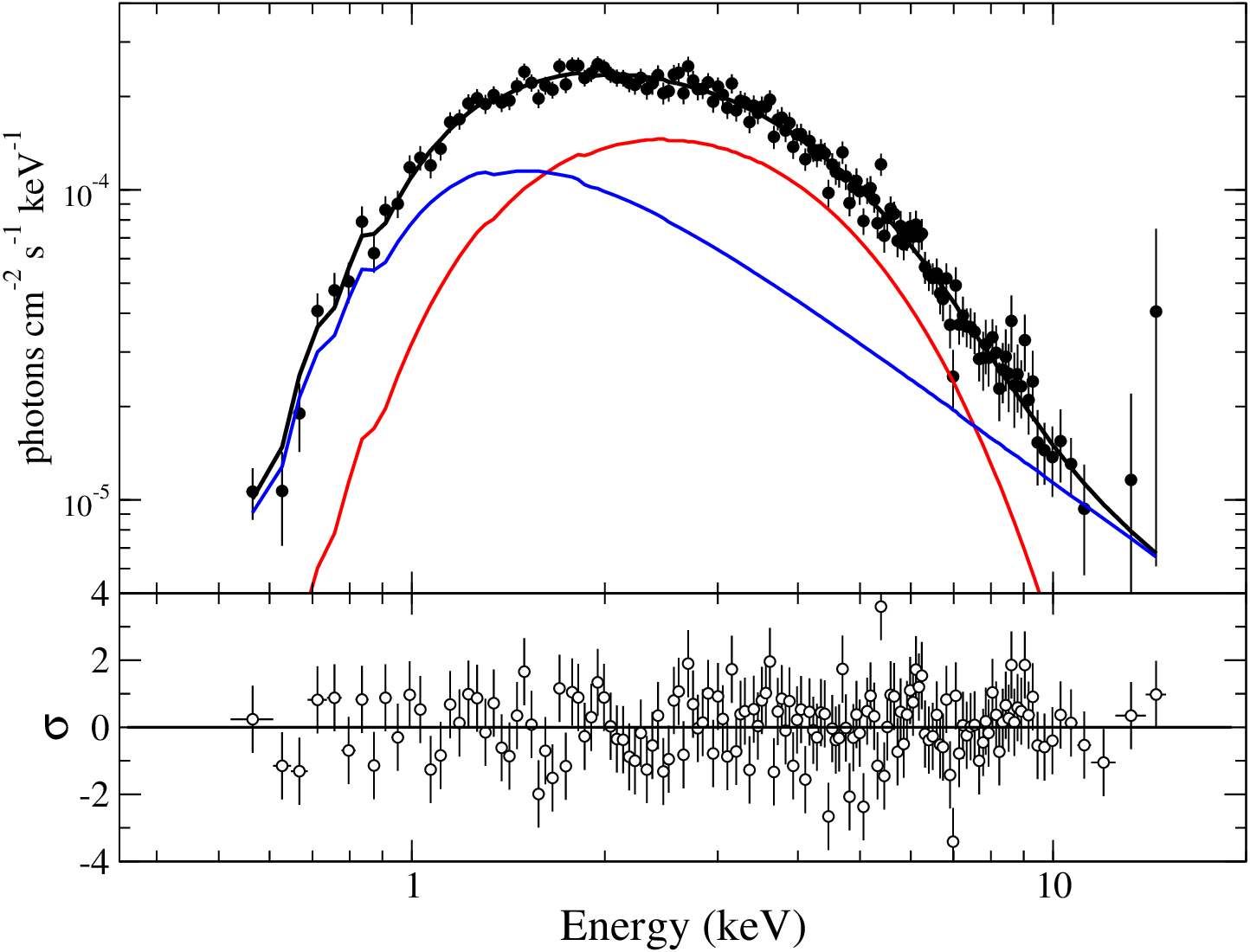} } 
\caption[]{PN spectrum of \igr\ (circles) and best-fit model
(black line), which consists of a blackbody (red line) and a power law 
(blue line).  The bottom panel gives the residuals in terms of sigmas with 
error bars of size one.}
\label{pn-spec}
\end{figure}

\subsection{Spectral analysis}

We extracted a PN energy spectrum using the same source and background regions
as for the timing analysis and filtering criteria described. For the spectral
analysis we further cleaned the data by accepting only the good times when sky
background was low. We rebinned the energy spectra by requiring at least 25
counts for each energy bin. 

Single-component models do not give good fits. An absorbed blackbody model
leaves residuals with a sinusoidal structure and differences between the data
and the model of $\simmore 2\sigma$ in 17\% of the bins. This model gives a
reduced $\chi^2=2$ for 150 degrees of freedom (dof). A single power law does not
describe the spectrum either with $\chi^2_{\rm red}= 4.5$ for 150 dof.  The
addition of a cutoff provides an acceptable fit ($\chi^2_{\rm red}= 1.07$ for
149 dof). The best-fit spectral parameters for the power-law plus cutoff model
are: $N_{\rm H}=(4.7\pm0.3) \times 10^{21}$ cm$^{-2}$, $\Gamma=-0.40\pm0.09$, and
$E_{\rm cut}=2.1\pm0.1$ keV. These values are significantly different from the
typical values in accreting pulsars and are hard to interpret. The photon index
and cutoff energy in accreting pulsars vary typically in the range $0.5-1.5$ keV
and $10-20$ keV, respectively. 

The combination of a blackbody and a power law provides an excellent fit with 
$\chi^2_{\rm red}= 1.01$ for 148 dof. In this case, the following best-fit
parameters are found:  $N_{\rm H}=(6.2\pm0.5) \times 10^{21}$ cm$^{-2}$,
$kT_{\rm bb}=1.16\pm0.03$ keV, and $\Gamma=1.5\pm0.1$.  The uncertainties
correspond to $1\sigma$ errors. The normalization of the blackbody component
agrees with an emitting region of $R_{\rm bb}=200\pm50$ m in size. The column
density was obtained assuming abundances given in \citet{anders89} and
cross-sections from \citet{verner96}. Figure~\ref{pn-spec} shows the EPIC-PN
spectrum and the two-component model described above. The total  absorbed
X-ray luminosity in the energy range 0.2--12 keV is $1.5\times 10^{34}$ \ergs\
($1.8\times 10^{34}$ \ergs, unabsorbed), assuming a distance of 4.5 kpc
\citep{reig10b}. The contribution from each model component to the total
luminosity is $L_{\rm bb}=9.3 \times 10^{33}$ \ergs\ and $L_{\rm pl}=8.6 \times
10^{33}$ \ergs. The contribution of the blackbody component to the source
luminosity amounts to $\sim$ 52\%. This value is somewhat larger but comparable
to that measured in other BeXBs
\citep{lapalombara06,lapalombara07,lapalombara09}. We found no evidence for an
iron emission line. The upper limit on the equivalent width of a narrow
($\sigma=0.1$ keV) 6.4 keV emission line is 30 eV.

\section{Discussion}

We have carried out an X-ray timing and spectral analysis of \igr\ at a time
when no X-ray emission should be expected. The circumstellar (decretion) disk
around the Be star's equator had presumably vanished at the time of the {\it
XMM-Netwon} observation. Not only we detected the source at a relatively high
level ($10^{34}$ \ergs), but also report X-ray pulsations for the first time in
this source. 

\subsection{Coupling between the periodic and aperiodic variability}
\label{coupling}

The power spectrum of \igr\ is characterised by the narrow peaks of the periodic
modulation and its harmonics and by a noise component whose power increases
toward lower frequencies. Red noise is a common characteristic in all kind of
accreting binaries.  It is attributed to instabilities in the accretion flow as
it approaches the compact object. If these instabilities are produced after the
flow is trapped by the magnetic field, they should be affected by the periodic 
modulation \citep{lazzati97,burderi97}. The power spectrum of \igr\ is very
similar to that of SMC X--1 \citep{burderi97}, Vela X--1 and 4U\,1145--62
\citep{lazzati97} and shows a strong coupling of the red noise with the periodic
modulation.   This coupling manifests observationally by the broadening of the
wings of the narrow peaks due to the periodic modulation and becomes more
apparent when the red-noise power increases short-ward of the pulsar frequency,
as it is the case in \igr.

\subsection{Origin of the quiescent X-ray emission}

In this section, we discuss the possible mechanisms that may
account for the X-ray emission of accreting pulsars at low luminosities. Accretion can explain the X-ray luminosity observed in X-ray pulsars in a large
range, from $10^{34}$ to $10^{38}$ \ergs. Two are the basic ingredients in the
accretion mechanism: a source of matter and a strong gravitational potential. In
accreting pulsars, the magnetic field also plays a fundamental role because
together with the mass accretion rate it defines the size of the magnetosphere. 
In BeXBs, the source of matter is the circumstellar disk around the Be star,
while the source of strong gravity is the neutron star.  

Accretion may cease if the mass accretion rate  $\dot{M}$ goes below a critical
value. When the radius of the magnetosphere, $r_{\rm m}$ grows beyond the
co-rotation radius, $r_{\rm co}$, (at which the angular velocity of Keplerian
motion is equal to that of the neutron star), the centrifugal force prevents
material from entering the magnetosphere. This is known as the propeller effect
\citep{illarionov75} or centrifugal inhibition of accretion \citep{stella86}.
While $r_{\rm co}$ does not depend on quantities that vary substantially with
the mass accretion rate, $r_{\rm co}=(GM_{\rm NS}P_{\rm spin}^2/4\pi)^{1/3}$,
the radius of the magnetosphere strongly depends on the mass accretion rate
$r_{\rm m}=(\mu^4/(2GM_{\rm NS} \dot{M}^2))^{1/7}$, where $\mu$ is the magnetic
moment. If $\dot{M}$ is highly reduced, then the inequality may reverse and
become $r_{\rm m} > r_{\rm co}$, in which case the propeller mechanism sets in.
Thus the propeller effect is expected to occur at the end of an X-ray outburst
or when there is no supply of matter.

By equating the co-rotation radius to the radius of the magnetosphere, a
minimum X-ray luminosity, corresponding to the minimum mass accretion rate, 
below which the propeller effect sets in can be determined 
\citep{stella86,campana02}

\begin{eqnarray}
\label{prop}
L_{\rm min}(R_{\rm NS}) = 3.9 \times 10^{37} \, k^{7/2}\,
B_{12}^2 \, P_{\rm spin}^{-7/3} \, M_{1.4}^{-2/3} \, R_6^5  \, \,  {\rm erg \, s^{-1}} 
\end{eqnarray}

\noindent where $M_{1.4}$ and $R_{6}$ are the mass and radius of the neutron
star in units of 1.4 $\msun$ and $10^6$ cm, respectively, $P_{\rm spin}$ the
spin period in seconds, $B_{12}$ the magnetic field in units of $10^{12}$ G, and
$k$ is the ratio of the magnetosphere to the Alfen radius (typically $k=0.5$ for
disk accretion). For typical values of the magnetic field (a few times $10^{12}$
G), $L_{\rm min}$  is of the order of $\sim 10^{36}$ erg s$^{-1}$ in rapid
rotating pulsars ($P_{\rm spin}\sim 1$ s). For long pulsing systems ($P_{\rm
spin}\sim 100$ s),  it is significantly lower,  
$L_{\rm min} \sim 10^{32}-10^{33}$ erg s$^{-1}$.

When the propeller effect is at work, X-rays may be generated through
magnetospheric accretion \citep{corbet96,campana01}. In this scenario, the
matter in the accretion flow is halted at the magnetospheric boundary and would
not reach the neutron star surface.  On first approximation, the accreted
luminosity will be driven by the same mass transfer as in the accretor regime
but with the magnetospheric radius replacing the neutron star radius
$L_{m}\approx GM_{\rm X}\dot{M}/r_m$. The maximum luminosity in this regime
will occur when the magnetospheric radius equals the corotation radius
\citep{campana02}

\begin{eqnarray}
\label{mag}
L_{\rm mag}(r_{\rm co})= L_{\rm min}(R_{\rm NS})\frac{R_{\rm NS}}{r_{\rm co}}
\end{eqnarray}

\noindent Although X-ray pulsations may still be produced in this state, the pulse
fraction should be lower than in the case of accretion onto the polar caps. 
The luminosity is expected to be very low $L_{\rm mag}\simless 10^{31}$
\ergs\ for systems with $P_{\rm spin}> 100$ s.

Because BeXBs contain massive companions, one may expect that matter could still
proceed from the optical star into the compact object via the stellar wind.
To estimate the X-ray luminosity that results from direct accretion from a
stellar wind, we assume that all the gravitational energy is converted into X-rays,  $L_{\rm
wind}=GM_{\rm X}\,\dot{M}_{\rm acc}/R_{\rm X}$ and that the fraction of the
stellar wind captured by the neutron star is \citep{frank02}

\begin{eqnarray}
\frac{\dot{M}_{\rm acc}}{\dot{M}}=\frac{G^2\,M_{\rm NS}}{a^2\,v_{\rm w}^4(a)}
\end{eqnarray}

\noindent where $a$ is the orbital separation, $\dot{M}$ the mass-loss rate of
the Be star, and $v_{\rm w}$ the terminal velocity of the stellar wind. Using
Kepler's law to replace $a$ we finally obtain \citep[see also][]{waters88}

\begin{eqnarray}
 L_{\rm wind} \approx 4.7 \times 10^{37} 
 M_{1.4}^3 \, R_6^{-1} \, M_{\rm *}^{-2/3} \, P_{\rm orb}^{-4/3} \, 
 \dot{M}_{-6} \, v_{\rm w,8}^{-4} \, \,
  {\rm erg \, s^{-1}}
\end{eqnarray}

\noindent where $M_*$ is the mass of the optical star in solar masses and
$P_{\rm orb}$ the orbital period of the binary in days. The Be star's mass-loss
rate is in units of $10^{-6}$ $\msun\, {\rm yr}^{-1}$ and the wind velocity in
units of $10^8$ cm s$^{-1}$. The classical supergiant X-ray binaries are
believed to be powered by accretion via a strong stellar wind.

The alternative to direct accretion from a high- or low-velocity outflow is that
accretion proceeds through magnetospheric leakage via an accretion disk
\citep[][see also \citealt{syunyaev77}]{tsygankov17a,tsygankov17b}. If the mass
accretion rate is low enough that the temperature outside the magnetosphere
remains below 6500 K, i.e. the temperature at which hydrogen recombines (above
6500 K hydrogen is ionised), then a cold disk is formed. Accretion is expected
to proceed at a low rate. The X-ray luminosity in this state is 
\citep{tsygankov17a}

\begin{eqnarray}
\label{coldisk}
L_{\rm disk} \approx 7 \times 10^{33} A ^{-7/13} k^{21/13} 
M_{1.4}^{3/13} \, R_6^{23/13} \, B_{12}^{12/13} \, T_{6500}^{28/13} \, \,
{\rm erg \, s^{-1}}
\end{eqnarray}

\noindent where $A$ is a parameter that depends on the the location at which the
viscous stress disappears as it interacts with the magnetosphere and varies in
the range 0.06--1 ($A\approx0.06$ if the stress disappears at $r_{\rm m}$).
According to this scenario, if  $L_{\rm min} > L_{\rm disk}$, then the pulsar
will transit to the propeller state. Otherwise, accretion at a low rate would
proceed from the cold disk. 


If accretion stops completely, then the only other mechanism that can also give
rise to X-rays is the cooling of the neutron star. During the accretion phase,
the crust of a neutron star is heated by nuclear reactions. This heat is
conducted inwards, heating the core, and outwards, where it is emitted as
thermal emission from the surface. After the  active accretion period, the crust
of a neutron star cools by X-ray emission until it reaches thermal equilibrium
with the core emission corresponding to the quiescent state
\citep{brown98,rutledge07,wijnands16}.

In summary, two are the basic mechanisms to explain the X-ray emission of
accreting pulsars at very low luminosities, namely accretion
and cooling of the neutron star surface. In either case, the physical conditions
on which these mechanisms occur lead to different scenarios. Accretion may
realise at the magnetosphere, via a stellar wind or through an accretion disk.
Likewise, the main contribution to the cooling process may come from the entire
surface of the neutron star or from the polar caps. In principle,  timing and
spectral analysis may help distinguish between the various possibilities. In
general terms, if accretion takes place, we expect a power-law (from
Comptonization) dominated spectrum and variability ($rms > 10$\%), while a
thermal spectrum and white noise dominated power spectrum is expected in the
case of cooling of the polar caps. A combination of the two cannot be ruled out.

\subsection{\igr}

In this section we discuss which of the possible scenarios describes best the
X-ray properties of \igr.  

First, we examine whether the system enters the propeller state, as expected by
the disappearance of the source that supplies matter to the neutron star.
Unfortunately, there is no direct estimate of the magnetic field of the neutron
star in \igr, hence $L_{\rm min}$ (eq.~(\ref{prop})) cannot be calculated. 
With a spin period of 373.2 s, we would expect $L_{\rm min}$ to be relatively
low. Assuming $B\sim 5 \times 10^{12}$, the luminosity at which the source would
enter the propeller state is $L_{\rm min} \sim 10^{33}$ erg s$^{-1}$. This is
well below the observed luminosity. Reversing the argument, if the observed
luminosity corresponded to the propeller state, then a magnetic field of $B\sim
2 \times 10^{13}$ G would be needed, which is three times larger than the
largest measured magnetic field in an accreting pulsar implied by the detection
of cyclotron lines \citep{yamamoto14,walter15}. If centrifugal inhibition of
accretion was at work in \igr, then the X-ray emission could in principle result
from magnetospheric accretion. However, there are several reasons why this
mechanism can be ruled out. When the source enters the centrifugally inhibited
state, a sharp drop of one or two orders of magnitude in luminosity (from
$L_{\rm min}$ to $L_{\rm mag}$) in the accreting luminosity should be observed.
This drop would occur on time scales of a few days
\citep{corbet96,campana02,tsygankov16,reig16b}. The X-ray luminosity expected
from magnetospheric accretion is far too low even in the case of a very strong
magnetic field, $L_{\rm mag} \sim 10^{30}$ \ergs (eq.~\ref{mag}).  This low
luminosity is a consequence of the strong dependence of $L_{\rm mag}$ on $P_{\rm
spin}$ ($L_{\rm mag} \propto P^{-3}_{\rm spin}$).   Second, the best-fit
spectral parameters imply a small emitting radius, $R_{\rm bb}\sim 0.2$ km,
which is too small to come from an extended region  at $R_{\rm m}\sim 10^9$ cm.
Finally, the coupling of the periodic signal and the noise indicates that
particles in the accretion flow have moved down along the magnetic field
lines close to the surface of the neutron star.

The cooling of the neutron star surface also encounters a number of problems as
the main mechanism producing the X-ray emission in \igr.  The blackbody
temperature and small emitting area imply that the cooling cannot occur over the
entire surface of the pulsar. In fact,  the large pulse fraction and broad pulse
profiles indicate that the emission arises from a rotating region which is
hotter (and more luminous) than the rest of the surface of the neutron star. 
Although non-uniform cooling can be attributed to the strong magnetic field
\citep{geppert06,wijnands16},  the X-ray spectrum of \igr\ deviates from pure
thermal emission. There is excess emission above $\sim$ 7 keV in the form of a
power law, indicating the presence of non-thermal processes. Another result that
argues against a purely thermal process in \igr\ is the degree of variability
and shape of the power spectral continuum.  The {\it XMM-Newton} power spectrum
shows strong low-frequency noise (red noise). White noise, i.e. no dependence of
power with frequency, would be expected in case of thermal cooling. Red noise is
believed to be produced by aperiodic variability associated with instabilities
in the accretion flow.  Finally, the observed X-ray luminosity is higher than
predicted from crustal heating. The X-ray luminosity expected from crustal
heating depends on the time-averaged accretion rate as $L_q\sim 6 \times
10^{32}(\dot{M}/1\times10^{-11})$ $\msun$ yr$^{-1}$, where $\dot{M}$ is the
average accretion rate including outbursts
\citep{brown98,rutledge07,wijnands13}. Typical values of $L_q$ in BeXBs are well
below $10^{34}$ \ergs\ \citep{tsygankov17b}.

Although crustal heating is not the dominating mechanism that accounts for the
observed X-ray emission in \igr, a strong thermal component is present. The high
temperature, small area, and large pulse fraction strongly suggest that the
origin of this component is the polar caps. This thermal component has been
observed in many, probably all,  accreting X-ray pulsars \citep{hickox04}. In
fact the values of $kT_{\rm bb}=1.16\pm0.03 $ keV and   $R_{\rm bb}=200\pm50$ m
agree well with those of 1\,A0536+262 \citep{mukherjee05}, RX J0146.9+6121
\citep{lapalombara06}, X-Per \citep{lapalombara07}, and RX\,J1037.5--5647
\citep{lapalombara09}. All these systems were observed at similar luminosity to
that of \igr\  ($L_X \simless 10^{35}$ \ergs) and displayed a similar spectrum,
namely an absorbed power law plus blackbody emission. An important difference of
\igr\ with respect to all these systems (except for RX\,J1037.5--5647 for which
no information could be found) is that at the time of the X-ray observation,
they all had \ha\ in emission, indicating the presence of the equatorial disk
\citep{reig16a}.

In summary, {\it i)} the amplitude of the X-ray variability ($rms=40$\%, red
noise), {\it ii)} the detection of pulsations with a high pulse fraction
($PF\simmore 50$\%), {\it iii)} the power-law component ($\Gamma=1.5$),  
{\it iv)} the coupling between periodic and aperiodic variability, and {\it v)}
the small emitting area strongly suggest that X-rays from \igr\ detected  during
the {\it XMM-Newton} observation result from accretion.  The question that
remains to be answered is how accretion could proceed when the main source of
matter was exhausted. 

Let us know examine the possibility of accretion from a stellar wind. Be stars,
by definition,  are dwarf, subgiant or giant objects (luminosity class III$-$V).
Consequently, the stellar wind is weak. Assuming typical parameters  $R_X=10^6$
cm, $M_{\rm X}=1.4\,\msun$, $M_{*}=15\, \msun$ (as expected for a B0.5V star
\citealt{reig10b}), $\dot{M_{\rm wind}} \sim 1 \times 10^{-8}$ $\msun$ yr$^{-1}$
\citep{prinja89,vink00}, $v_{\rm wind}\sim  v_{\rm esc}\approx 8.5 \times 10^8$ cm
s$^{-1}$, and $P_{\rm orb}=100$ days, the X-ray luminosity from the stellar wind
is well below $10^{33}$ \ergs.  $L_{\rm wind}$ approaches the observed X-ray
luminosity only if  $P_{\rm orb}\simless 10$ days  or  $v_{\rm wind}$ is
abnormally low ($400-500$ km s$^{-1}$). BeXBs have orbital periods well above 20
days. From the $P_{\rm orb}$--$P_{\rm spin}$ relationship
\citep{corbet86,reig11}, a system with $P_{\rm spin}=373$ s should have $P_{\rm
orb}\simmore 100$ days.  An exception is SAX\,J2103.5+4545, which is the BeXB
with the shortest orbital period of 12.7 days but has a pulse period of 358 s
\citep{baykal02}. Abnormally slow winds have been reported in one BeXB,
4U\,2206+54 \citep{ribo06}. However, 4U\,2206+54 is a rather peculiar object
\citep{blay06}.  \citet{reig14b} argued that a weak and small disk might be
present in the BeXB IGR\,J21343+4738 even though an absorption \ha\ profile was
observed. A low-velocity outflow from this highly debilitated disk with $v_{\rm
wind}\sim 300-500$ km s$^{-1}$ could  explain the X-ray luminosity of $\sim
10^{35}$ \ergs, measured in that system. Accretion from an invisible (in the
optical and IR bands) disk was also proposed by \citet{ikhsanov01a} to explain the
quiescent emission of 1A\,0535+262. We conclude that direct accretion from
stellar wind cannot explain the origin of the X-ray emission in \igr, although a
slow outflow from a weak optically undetectable equatorial disk cannot be ruled
out. The fact that the \ha\ line showed clear evidence for emission less than
two months after the {\it XMM-Netwon} observation and that the \ew\ was slightly
smaller than that expected from a fully photospheric line possibly indicate the
presence of a very weak residual disk.

The alternative to direct accretion is the formation of an accretion disk. The
centrifugal barrier caused by the rotating magnetosphere is strongly suppressed
if an accretion disk is present. Such a disk will form if  the source luminosity
is below the limiting luminosity given in Eq.~(\ref{coldisk}).   If this
luminosity is reached before the propeller minimum luminosity 
the source will transit to stable accretion from an entirely recombined cold
disk \citep{tsygankov17a}.  For a typical X-ray pulsar with a magnetic field
around $2\times10^{12}$ G,  $L_{\rm disk}\sim 2 \times 10^{34}$ erg s$^{-1}$
\ergs, while $L_{\rm min}$ would be two orders of magnitude lower. Hence
$L_{\rm min} < L_{\rm disk}$, and the condition under which the source may
accrete matter from the cold disk is met. 

The long spin period of \igr\ and the detection of X-ray pulsations put this
source along with 1A 0535+26, 4U 1145--619, and 1A 1118--615, in a category of
systems with accretion powered quiescent emission.

\section{Conclusion}

We perform an X-ray timing and spectral analysis of the BeXB candidate \igr. We
discovered pulsations with a pulse period of 373.2 s. Hence, our observations
convert \igr\ from a candidate BeXB to a confirmed accreting pulsar.  We also
found strong coupling between the periodic and aperiodic variability, which
manifests as a broadening of the base of the pulse peak in the power spectrum.
The X-ray spectrum is well described by a thermal (blackbody) and a non-thermal
(power law) components, affected by interstellar absorption. We did not find
evidence for iron emission at 6.4 keV. The presence of the non-thermal component
(interpreted as coming from bulk Comptonization), the strong red noise in the
power spectrum, the pulsations, and the small emitting area of the thermal
component (polars caps)  are all considered as signatures of accretion and rules
out cooling of the accretion-heated neutron star crust as the sole mechanism
that explains our {\it XMM-Newton} observation. Although accretion powers the
quiescent emission in \igr, we cannot tell whether the source of accreted matter
is a slow wind coming out from a weak decretion disk or from a recombined cold
accretion disk. We should emphasise that the order of magnitude calculations
presented in this work strongly depend on the magnetic fields strength. In the
absence of a direct measurement, for example, through the detection of a
cyclotron resonant scattering feature, these numbers should be taken with
caution.

\begin{acknowledgements}

We thank N. Schartel for his fast response to our request to observe the source
and for approving a TOO observation under Discretionary time. Skinakas
Observatory is run by the University of Crete and the Foundation for Research
and Technology-Hellas. A.Z. acknowledges funding from the European Union's
Seventh Framework Programme (FP/2007-2013)/ERC Grant Agreement No. 617001.

\end{acknowledgements}

\bibliographystyle{aa}
\bibliography{../../../artBex_bib}

\end{document}